# Terahertz spectroscopy for all-optical spintronic characterization of the spin-Hall-effect metals Pt, W and Cu$_{80}$Ir$_{20}$

T. S. Seifert[1,2,3], N.M. Tran[4], O. Gueckstock[1], S.M. Rouzegar[1], L. Nadvornik[1,2], S. Jaiswal[5,6], G. Jakob[5,7], V.V. Temnov[4], M. Münzenberg[8], M. Wolf[1], M. Kläui[5,7], T. Kampfrath[1,2*]

1. Department of Physical Chemistry, Fritz Haber Institute of the Max Planck Society, 14195 Berlin, Germany
2. Department of Physics, Freie Universität Berlin, 14195 Berlin, Germany
3. Department of Materials, Eidgenössische Technische Hochschule Zürich, 8093 Zürich, Switzerland
4. Institut des Molécules et Matériaux du Mans, UMR CNRS 6283, Le Mans Université, 72085 Le Mans, France
5. Institute of Physics, Johannes Gutenberg University, 55128 Mainz, Germany
6. Singulus Technologies AG, 63796 Kahl am Main, Germany
7. Graduate School of Excellence Materials Science in Mainz, 55128 Mainz, Germany
8. Institute of Physics, Ernst Moritz Arndt University, 17489 Greifswald, Germany

* Email: tobias.kampfrath@fu-berlin.de

**Abstract.** Identifying materials with an efficient spin-to-charge conversion is crucial for future spintronic applications. In this respect, the spin Hall effect is a central mechanism as it allows for the interconversion of spin and charge currents. Spintronic material research aims at maximizing its efficiency, quantified by the spin Hall angle $\Theta_{SH}$ and the spin-current relaxation length $\lambda_{rel}$. We develop an all-optical contact-free method with large sample throughput that allows us to extract $\Theta_{SH}$ and $\lambda_{rel}$. Employing terahertz spectroscopy and an analytical model, magnetic metallic heterostructures involving Pt, W and Cu$_{80}$Ir$_{20}$ are characterized in terms of their optical and spintronic properties. The validity of our analytical model is confirmed by the good agreement with literature DC values. For the samples considered here, we find indications that the interface plays a minor role for the spin-current transmission. Our findings establish terahertz emission spectroscopy as a reliable tool complementing the spintronics workbench.

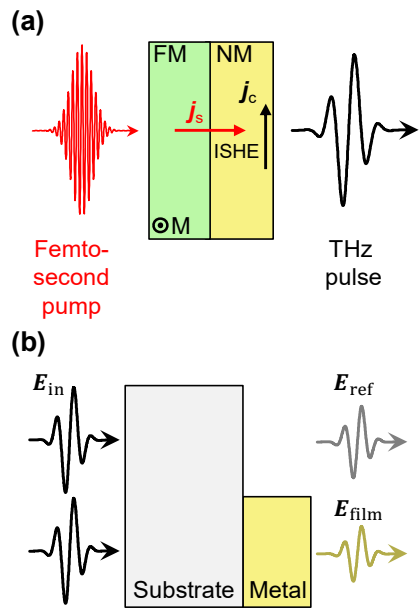

**Figure 1. Schematic of the experiment. (a)** Terahertz emission experiment. A femtosecond near-infrared pump pulse excites electrons in both the ferromagnetic (FM, in-plane magnetization $M$) and non-magnetic (NM) metal layer. Due to the asymmetry of the heterostructure, a spin current $j_s$ is injected from the FM into the NM material where it is converted into an in-plane charge current $j_c$ by the inverse spin Hall effect (ISHE). The sub-picosecond charge-current burst leads to the emission of a terahertz (THz) pulse into the optical far-field. **(b)** Terahertz transmission experiment. A THz transient $E_{in}$ is incident onto either the bare substrate or onto the substrate coated by a thin metal film. By comparing the two transmitted waveforms $E_{ref}$ and $E_{film}$, the metal conductivity at THz

frequencies is determined.

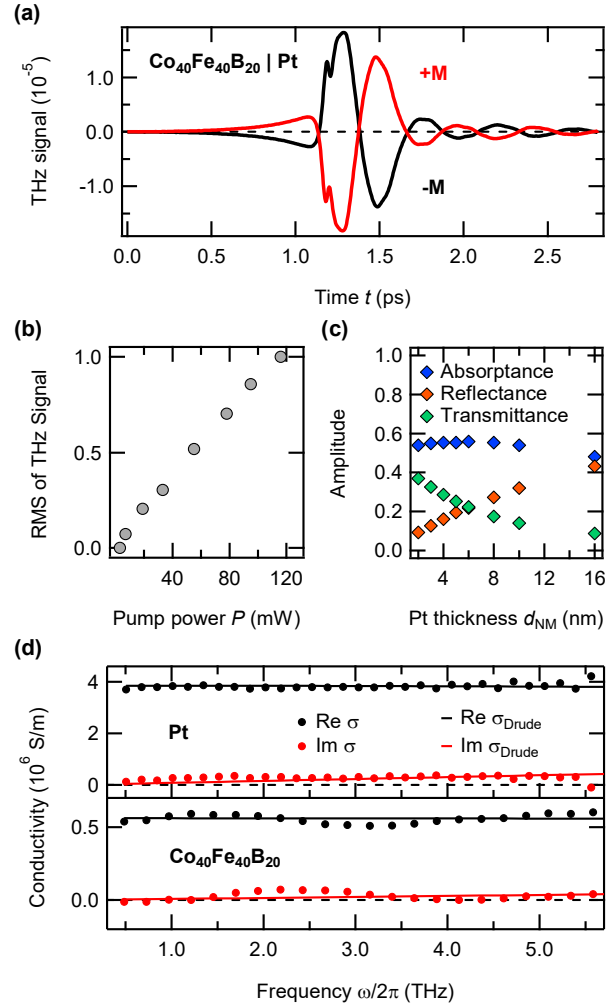

**Figure 2. Typical THz emission raw data and sample characterization. (a)** THz emission signal measured from a $C_{40}F_{40}B_{20}$(3 nm)|Pt(3 nm) bilayer for two opposite orientations of the sample magnetization ($\pm M$). **(b)** Normalized pump-power dependence of the THz signal amplitude (RMS) for one orientation of the sample magnetization. **(c)** Pump-light absorptance, transmittance and reflectance as function of the Pt-layer thickness. **(d)** Frequency-dependent THz conductivities measured by THz transmission experiments (black and red dots) along with fits obtained by the Drude model (black and red solid lines).

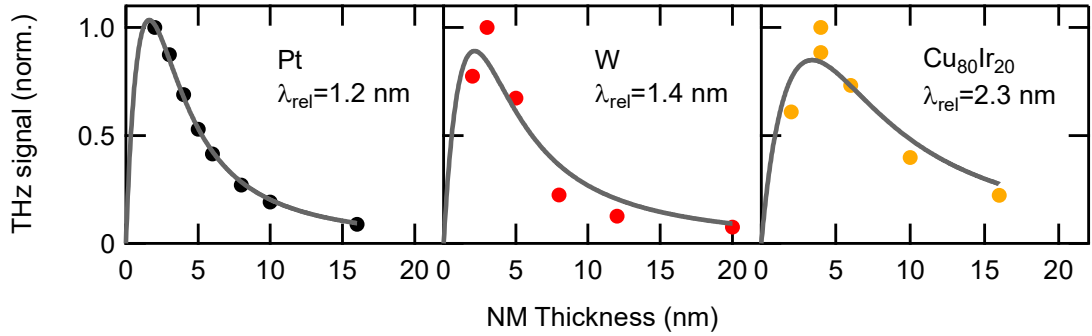

**Figure 3. Thickness dependence of the THz-emission signal.** THz-signal amplitude (RMS) as a function of the NM layer thickness divided by the thickness-dependent pump absorptance for $Co_{40}Fe_{40}B_{20}(3\,nm)|Pt$ (markers in left panel), $Co_{20}Fe_{60}B_{20}(3\,nm)|W$ (center panel), and $Fe(3\,nm)|Cu_{80}Ir_{20}$ (right panel). Grey solid lines show fits based on Eqs. (1) and (2) with the relaxation length $\lambda_{rel}$ and a global amplitude as free parameters. We obtain $\lambda_{rel} = (1.2 \pm 0.1)\,nm$, $(1.4 \pm 0.5)\,nm$ and $(2.3 \pm 0.7)\,nm$ for Pt, W and $Cu_{80}Ir_{20}$, respectively. For each NM material, the experimental data is normalized by the maximum amplitude.

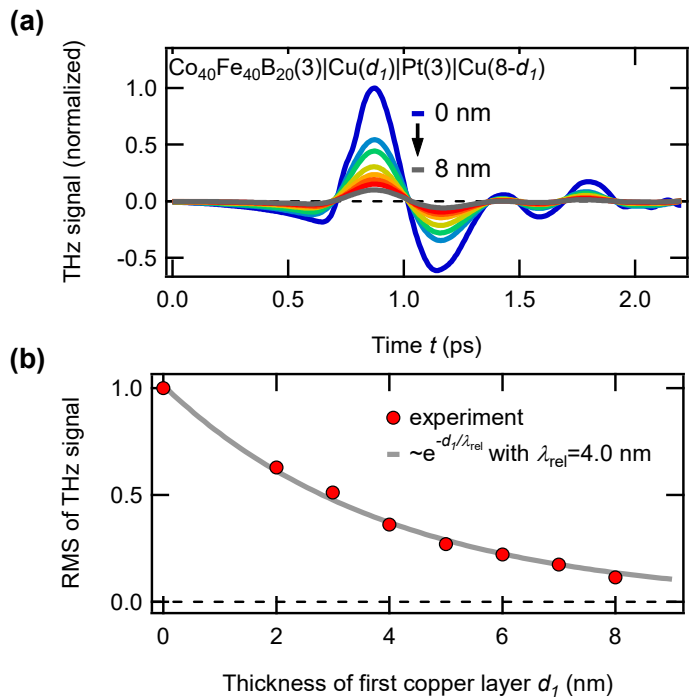

**Figure 4. Role of the interface in THz emission studies. (a)** THz waveforms emitted from $Co_{40}Fe_{40}B_{20}|Cu(d_1)|Pt|Cu(8\,nm-d_1)$ structures with fixed $Co_{40}Fe_{40}B_{20}$ and Pt thicknesses of 3 nm and $d_1$ ranging from 0 to 8 nm. **(b)** THz signal amplitudes (RMS) vs $d_1$ together with a single exponential fit $\propto \exp(-d_1/\lambda_{rel})$ with $\lambda_{rel} = (4.0 \pm 0.1)\,nm$. The experimental data is normalized by the maximum signal amplitude.

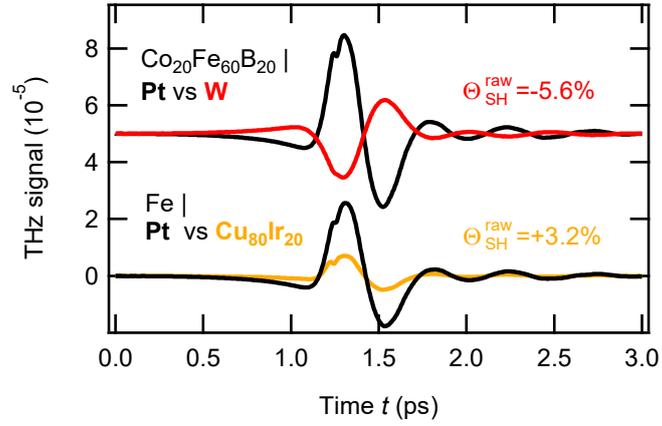

**Figure 5. Comparison of raw THz-emission data from Pt, W, and $Cu_{80}Ir_{20}$.** THz waveforms emitted from FM|NM structures where NM is W (red) and $Cu_{80}Ir_{20}$ (orange) in comparison to a FM|Pt reference structure with the same layer thicknesses (black curves). The values of $\Theta_{SH}^{raw}$ indicate the spin Hall angles of the respective NM materials estimated by calculating the ratios of the bare THz-signal amplitudes (RMS) and assuming a spin Hall angle of 12.0% for the Pt reference.

**Table 1. Overview of samples used for THz-emission measurements together with the corresponding reference samples.** Numbers in parentheses indicate the film thickness in nm. In the last row, $d_1$ ranges from 0 to 8 nm. The substrate thickness is 500 µm.

| NM material | Sample structure | Reference sample structure |
|---|---|---|
| Pt | Glass\|\|$Co_{40}Fe_{40}B_{20}$(3)\|Pt(2-16) | - |
| W | Glass\|\|$Co_{20}Fe_{60}B_{20}$(3)\|W(2-20) | Glass\|\|$Co_{20}Fe_{60}B_{20}$(3)\|Pt(3) |
| $Cu_{80}Ir_{20}$ | Glass\|\|Fe(3)\|$Cu_{80}Ir_{20}$(2-16)\|$AlO_x$(3) | Glass\|\|Fe(3)\|Pt(4) |
| Cu | Glass\|\|$Co_{40}Fe_{40}B_{20}$(3)\|Cu($d_1$)\|Pt(3)\|Cu(8-$d_1$) | - |

**Table 2. Sample details for THz-transmission measurements.** Drude-model fit parameters $\sigma_{\mathrm{DC}}$ and $\Gamma$, conductivity literature values $\sigma_{\mathrm{DC}}^{\mathrm{lit}}$, and substrate-thickness variation $\Delta d_{\mathrm{sub}}$. Numbers in parentheses indicate the film thickness in nm. The substrate thickness is 500 μm.

| Material | Sample | $\sigma_{\mathrm{DC}}$ ($10^6$ S/m) | $\sigma_{\mathrm{DC}}^{\mathrm{lit}}$ ($10^6$ S/m) | Ref. | $\Gamma$ ($2\pi \cdot 10^{12}$ Hz) | $\Delta d_{\mathrm{sub}}$ (μm) |
|---|---|---|---|---|---|---|
| $Co_{40}Fe_{40}B_{20}$ | Glass\|\|$Co_{40}Fe_{40}B_{20}$(3)\|MgO(4) | 0.6 | 0.7 | [1] | 81 | -0.1 |
| Pt | Glass\|\|Pt(10) | 3.9 | 5.4 | [4] | 51 | 5.8 |
| $Co_{20}Fe_{60}B_{20}$ | Glass\|\|$Co_{20}Fe_{60}B_{20}$(6) | 0.6 | - | - | 25 | -7.3 |
| W | Glass\|\|W(10) | 1.9 | 2.0 | [2] | >100 | -5.9 |
| Fe | MgO\|\|Fe(3)\|AlO$_x$(3) | 2.7 | 4.0 | [3] | >100 | -2.2 |
| $Cu_{80}Ir_{20}$ | Glass\|\|$Cu_{80}Ir_{20}$(4)\|AlO$_x$(3) | 1.0 | 2.0 | [39] | 64 | -0.6 |

**Table 3. Spin-Hall parameters.** Spin-current relaxation length $\lambda_{\mathrm{rel}}$ along with the literature value $\lambda_{\mathrm{rel}}^{\mathrm{lit}}$. The table also provides the ratio of the bare THz emission signals $\Theta_{\mathrm{SH}}^{\mathrm{raw}}$ between a FM|NM heterostructure and the corresponding FM|Pt reference sample as well as the relative spin Hall angle $\Theta_{\mathrm{SH}}$ extracted by our analysis correcting for sample specific parameters [see Eqs.(1) and (2)] along with the corresponding literature value $\Theta_{\mathrm{SH}}^{\mathrm{lit}}$.

| NM Material | $\lambda_{\mathrm{rel}}$ (nm) | $\lambda_{\mathrm{rel}}^{\mathrm{lit}}$ (nm) | Ref. | $\Theta_{\mathrm{SH}}^{\mathrm{raw}}$ ($10^{-2}$) | $\Theta_{\mathrm{SH}}$ ($10^{-2}$) | $\Theta_{\mathrm{SH}}^{\mathrm{lit}}$ ($10^{-2}$) | Ref. |
|---|---|---|---|---|---|---|---|
| Pt | 1.2 | 1.1 | [4] | - | - | 12.0 | [40] |
| W | 1.4 | <3 | [17] | -5.6 | -4.0 | -4 | [17] |
| $Cu_{80}Ir_{20}$ | 2.3 | <5 | [38] | 3.2 | 1.9 | 2.1 | [38] |

# 1. Introduction

Using the electron spin to replace some parts of conventional electronics based on charge dynamics is the goal of spintronics research[5]. Its large potential for future information technology is highlighted by recently developed competitive spintronic applications such as spin-transfer- and spin-orbit-torque magnetic random-access memory[6,7,8]. Driven by fundamental and material-science research at the same time, future progress in spintronics crucially relies on the identification of materials with efficient spin-to-charge-current conversion.

Central effects in this regard are the spin Hall effect and its inverse (ISHE) which allow for an interconversion of spin and charge currents[9,10]. The efficiency of this process can be quantified by the product of the spin Hall angle $\Theta_{SH}$ and the spin-current relaxation length $\lambda_{rel}$. To identify materials with maximum $\Theta_{SH}$ and $\lambda_{rel}$, one needs to be able to screen many material candidates efficiently and accordingly optimize spintronic device parameters. To this end, a fast, reliable and straightforward characterization method is thus a key to realize future spin-based applications.

So far, most works relied on electrical measurements which require microstructuring and electrical contacting of the samples[11,12,13,14]. This procedure is costly, time-consuming and may complicate data interpretation due to additional contact resistance[15]. Other established techniques such as spin-Seebeck-effect measurements do not require microstructuring but still rely on electrical contacting[39]. Recently, non-contact methods have been developed which are based on inductive coupling between a microwave waveguide and the spintronic sample[16] or on magnetooptically detected ferromagnetic resonance[17].

Here, we present an all-optical scheme using terahertz (THz) emission spectroscopy (TES) of metallic heterostructures together with a detailed analysis of how to extract the key quantities $\Theta_{SH}$ and $\lambda_{rel}$. The operational principle is shown in Fig.1(a): The inversion-asymmetric sample is excited by a femtosecond near-infrared pump pulse, thereby driving an ultrafast spin current from the FM into the NM layer. This process can be considered as an ultrafast version of the spin-dependent Seebeck effect. In the NM layer, the ISHE converts the longitudinal spin flow into a transverse, in-plane charge current[18,19,20,21]. The sub-picosecond transverse charge-current burst leads to the emission of a THz electromagnetic pulse into the optical far-field where it is detected by electrooptic sampling[22,23]. This scenario can also be considered as a realization of the spin-galvanic effect[24].

It is worth noting that THz emission by spintronic multilayers consisting of nanometer-thick ferromagnetic (FM) and nonmagnetic (NM) metals has led to the development of efficient THz emitters[19,25,26,27,28,29] which are capable of providing a broadband and continuous spectrum from 1 to 30 THz without gaps (for details see ref. 19). In addition, we recently found agreement between the amplitude of the THz field emitted from FM|NM bilayers and the intrinsic spin-Hall conductivity of the employed NM material[19]. This observation is, however, surprising because $\Theta_{SH}$ and $\lambda_{rel}$ can strongly depend on the electron energy[30,31]. Moreover, the spin-current-injection efficiency is material- and interface-dependent[32,33]. These intriguing issues require a more detailed study to establish TES of magnetic heterostructures[34,35,36,37] as a reliable spintronic characterization tool. In the following, we focus on two familiar spintronic metals, namely platinum (Pt) and tungsten (W), as well as a new and promising copper-iridium alloy $Cu_{80}Ir_{20}$ (refs. 38,39).

# 2. Theoretical background

In analogy to DC-spin-pumping experiments[12], we describe the in-plane sheet charge-current density $J_c$ in the NM layer by a simple model which yields[19,20]

$$J_c = \frac{AF_{inc}}{d} j_s^0 \cdot \lambda_{rel} \tanh \frac{d_{NM}}{2\lambda_{rel}} \cdot \Theta_{SH}. \qquad (1)$$

Here, $A$ denotes the absorbed fraction of the incident pump-pulse fluence $F_{\text{inc}}$ (the latter being constant for all measurements throughout this work), $d$ is the entire metal thickness, $j_s^0$ is the spin-current density (per pump-pulse excitation density) injected through the FM/NM interface into the NM layer, $d_{\text{NM}}$ is the NM layer thickness, $\lambda_{\text{rel}}$ is the relaxation length of the ultrafast spin-current, and $\Theta_{\text{SH}}$ is the spin Hall angle.

Equation (1) accounts for the photoinduced injection of the spin current into the NM layer (first term), the multiple spin-current reflections in the NM layer (second term) and the ISHE (last term). For simplicity, we assume a spin-current reflection amplitude of unity at the NM/air and NM/FM interfaces. Moreover, the spin- and energy-dependent spin-current transmission[32,33] of the FM/NM interface is assumed to be approximately independent of the NM material. This assumption is supported by our previous work in which a strong correlation between THz signal amplitude and the spin Hall conductivity of the NM layer was found (see Fig. 2 in Ref. 19). Since the currents are driven by linear absorption of pump photons [see Fig. 2(b)], we assume that the sheet charge-current density scales linearly with the pump-pulse excitation density $A/d$.

Finally, to determine the resulting THz electric field, we assume a normally incident plane-wave-like pump pulse and a sufficiently thin metal film (total metal thickness $d$ minor compared to attenuation length and wavelength of the THz field within the metal). Under these assumptions, the THz electric field directly behind the sample reads[19,20]

$$E_{\text{THz}}(\omega) = \frac{eZ_0}{n_1 + n_2 + Z_0 \int_0^d dz\, \sigma(\omega, z)} J_c(\omega). \qquad (2)$$

Here, $n_1$ and $n_2 \approx 1$ are the refractive indices of substrate and air, respectively, $-e$ denotes the electron charge, $Z_0 \approx 377\,\Omega$ is the vacuum impedance, and $\sigma$ is the $z$-dependent metal conductivity at THz frequencies.

It is important to emphasize that the derivation of Eq. (2) relies on a collimated pump beam with a diameter much larger than all THz wavelengths considered. Therefore, the THz beam behind the sample can be approximated by a plane wave. This condition is not fulfilled in our experiment (see below) where the pump focus is substantially smaller than the detected THz wavelengths, which easily exceed 100 μm. However, as shown in the Supplementary Material, Eq. (2) remains valid within a relative error of less than 3% because the apertures of optical elements after the sample reduce the divergence angle of the THz beam to less than 20° with respect to the optical axis.

Equations (1) and (2) provide the key to determine $\Theta_{\text{SH}}$ and $\lambda_{\text{rel}}$ and naturally take us to the following three-step procedure: First, we characterize the involved metals in terms of their parameters unrelated to spintronics, namely $A$ and $\sigma$ [see Fig. 1(b)]. Second, we extract $\lambda_{\text{rel}}$ by varying the NM layer thickness in a THz emission experiment [see Fig. 1(a)]. Third, we address the spin-current transmission of the FM/NM interface. Eventually, the relative spin Hall angle $\Theta_{\text{SH}}$ is determined by a reference measurement using the widely studied metal Pt as the NM layer[40].

## 3. Experimental details

We employ femtosecond near-infrared laser pulses (central wavelength of 800 nm, duration of 10 fs, energy of 2.5 nJ, repetition rate of 80 MHz) from a Ti:sapphire laser oscillator to excite the sample under study [see Fig. 1(a)]. In the THz emission experiment [Fig. 1(a)], the pump beam is focused onto the sample (spot size of 22 μm full width of the intensity maximum). The emitted THz pulse is detected by electrooptic (EO) sampling[22,23] in a standard 1-mm-thick ZnTe crystal using a copropagating femtosecond near-infrared pulse derived from the same laser. To saturate the sample magnetization in the sample plane, an external field of 10 mT is applied. All measurements are carried out in a dry nitrogen atmosphere.

For the THz conductivity measurements[41] [Fig. 1(b)], the femtosecond pump pulses are replaced by broadband THz pulses generated in a spintronic THz emitter[19]. We measure the transmission of the THz pulse through the metallic sample film on the substrate and through a part of the substrate free of any metal [see Fig. 1(b) and Supplementary Material for details]. For these THz transmission experiments, we employ a 250-μm-thick GaP EO crystal as detector.

Note that the electrooptic signal of the THz pulse $S(t)$ is in general not simply proportional to the transient THz electric field $E_{\mathrm{THz}}(t)$ [Fig. 1(a)]. Instead, it is a convolution of $E_{\mathrm{THz}}(t)$ with the setup response function which captures the propagation and electrooptic sampling[42,43]. One can, in principle, retrieve the THz field $E_{\mathrm{THz}}(t)$ from the electrooptic signal $S(t)$ to eventually determine the dynamics of the charge current flowing inside the sample[43]. Throughout this study, however, all measurements are carried out under similar experimental conditions, thereby enabling a direct comparison of the measured raw data.

Tables 1 and 2 summarize the stacking structure of the samples used for TES and for THz transmission measurements, respectively. Details on the sample preparation can be found in the Supplementary Material.

**4. Results**

*4.1. Typical THz-emission raw data*

Figure 2(a) shows typical THz-emission raw data measured from a $C_{40}F_{40}B_{20}$(3 nm)|Pt(3 nm) bilayer. The signal reverses almost entirely when the sample magnetization is reversed, thus confirming its magnetic origin. For all studied samples, the signal contribution being even in the sample magnetization is found to be below 5% of the signal odd in the sample magnetization. It might originate from sample imperfections such as pinned magnetic domains[44], but is not considered further as the odd signal clearly dominates. Consequently, we concentrate only on the THz signal for one orientation of sample magnetization.

We observe that the THz-signal amplitude [root mean square (RMS) of the time-domain THz signal] increases linearly with pump fluence. This finding shows that a second-order nonlinear process in the pump laser field is operative [see Figs. 2(b) and S1], consistent with the linear scaling of the THz sheet charge-current density with $F_{\mathrm{inc}}$ as assumed in our model [see Eq. (1)].

We emphasize that the THz waveforms emitted from all samples exhibit similar temporal dynamics. Thus, a direct comparison of signal amplitudes (RMS) is possible.

*4.2. Optical and THz characterization*

To isolate the intrinsic spin-orbit-interaction-related parameters from the THz-emission signal, we first measure the optical absorptance of the pump pulse [see Fig. 2(c)]. We note that currents arising from a pump-induced gradient can be neglected for thicknesses smaller than 20 nm because the larger penetration depth of the pump electric field and its multiple reflections result in a constant electric field along the $z$-direction in the sample[19].

Second, we characterize the samples by THz-transmission experiments[41]. In the thin-film limit, the THz transmission can directly be related to the metal-film conductivity, which we extract in the frequency interval from 0.5 to 5.5 THz. Details of the extraction procedure are described in the Supplementary Material. Note that we treat the multilayers as a parallel connection of their constituents. Therefore, the total conductance of the multilayer is given by the sum of the individual layer conductances.

We find that all metal conductivities are well described by the Drude model [see Figs. 2(d), S3 and the Supplementary Material for details]. The DC conductivity $\sigma_{\mathrm{DC}}$ and electronic current relaxation rate $\Gamma$ are used as fit parameters whose values for all involved materials are given in Table 2. Note that a slight variation in the substrate thickness $\Delta d_{\mathrm{sub}}$ between sample and reference THz transmission measurements is accounted for by a phase-shift term[45,46] of the form $\exp[-i\omega(n_2 - 1)\Delta d_{\mathrm{sub}}/c]$ with the free parameter $\Delta d_{\mathrm{sub}}$ (see Table 2).

In the following, we consider the frequency-averaged THz conductivity in which each frequency is weighted by the relative spectral amplitude of the detected THz signal. A comparison of the extracted values of $\sigma_{\mathrm{DC}}$ and literature values $\sigma_{\mathrm{DC}}^{\mathrm{lit}}$ reveals a good agreement (see Table 2). Deviations, such as those in the cases of Fe and $Cu_{80}Ir_{20}$, are most likely caused by different thicknesses[47] or crystallinity of the films. With this procedure, we have determined all parameters unrelated to the spin-to-charge conversion in the NM layer [see Eqs. (1) and (2)].

*4.3. Thickness dependence*

The first spintronic key parameter to be extracted is the relaxation length $\lambda_{\mathrm{rel}}$ of the ultrafast spin current in the NM layer. To this end, we carry out measurements (see Fig. 3) in which the FM layer thickness is kept constant and the NM layer thickness is varied. The measured data are divided by the thickness-dependent pump-pulse absorptance and normalized by the maximum signal amplitude [see Figs. 2(c) and S2]. A least-squares fit using Eqs. (1) and (2) with the only free parameters $\lambda_{\mathrm{rel}}$ and a global amplitude, yields a $\lambda_{\mathrm{rel}}$ of $(1.2 \pm 0.1)$ nm for Pt, $(1.4 \pm 0.5)$ nm for W, and $(2.3 \pm 0.7)$ nm for $Cu_{80}Ir_{20}$ (see Fig. 3 and Table 3). The reduced chi-squared values of the fits are 1.11, 1.04 and 1.02, respectively. We find that the values of $\lambda_{\mathrm{rel}}$ are quite insensitive to the value of the spin-current reflection amplitude at the FM/NM interface, thereby underlining the robustness of our analysis.

The $\lambda_{\mathrm{rel}}$ values inferred here are consistent with previous THz-emission studies on FM|Pt bilayers[19]. They agree very well with literature values from DC experiments (see Table 3). The agreement is surprising as the DC experiments measure the DC spin diffusion length which does not necessarily coincide with the relaxation length of the THz spin current. The latter might be even shorter since spin flips usually occur only in a fraction of all electronic scattering events[48]. Furthermore, we find an inverse proportionality between $\sigma_{\mathrm{DC}}$ (see Table 2) and $\lambda_{\mathrm{rel}}$ (see Table 3). However, no correlation between the Drude scattering rate $\Gamma$ and the relaxation length $\lambda_{\mathrm{rel}}$ is seen. These interesting observations will be addressed in forthcoming studies.

*4.4. Role of the FM/NM interface*

To gain insights into the role of the interfacial spin-current transmission in the THz emission experiment, we study samples with an intermediate layer with varying thickness. As the material, we choose copper because it exhibits a negligibly small SHE[32] and efficient spin current transport[49] that can be employed for spin-switching devices[50].

In detail, we measure the emitted THz signal from $Co_{40}Fe_{40}B_{20}$(3 nm)|Cu($d_1$)|Pt(3 nm)|Cu(8 nm$-d_1$) stacks (see Table 1) with $d_1$ ranging from 0 to 8 nm [see Fig. 4(a)]. This choice of layer structure results in metal stacks of the same thickness and, thus, a pump-pulse absorptance of $A \approx 0.6$ and a THz outcoupling efficiency [Eq. (2)] which are both largely independent of $d_1$. Note that the Pt thickness of 3 nm is sufficient to prevent any sizeable spin-current transmission into the second copper layer (see Fig. 3).

Therefore, the raw data of Fig. 4(a) become directly comparable and provide a measure of the spin-current amplitude injected into the Pt "detector layer". Figure 4(b) shows that the THz-signal strength (RMS) decreases with increasing $d_1$. The experimental data can be very well fit (reduced chi-square value of 1.02) by a monoexponential $\propto \exp(-d_1/\lambda_{rel})$ with $\lambda_{rel} = (4.0 \pm 0.1)$ nm [see Fig. 4(b)]. We note that on one hand, this value of the THz-spin-current relaxation length in Cu is substantially larger than for metals with a sizeable SHE (see Fig. 3). On the other hand, it is small compared to DC spin diffusion lengths, which are on the order of 100 nm in Cu (ref. 51).

Interestingly, we do not observe indications of a discontinuity around $d_1 = 0$ nm. The continuous behavior suggests similar spin-current transmission efficiencies through $Co_{40}Fe_{40}B_{20}$/Pt compared to $Co_{40}Fe_{40}B_{20}$/Cu and Cu/Pt interfaces. It must be emphasized that this observation cannot be generalized and may, in contrast, strongly depend on materials and probe frequency. For example, recent gigahertz spin-pumping experiments on Co|Cu|Pt revealed a strong interface dependence of the spin-current transmission[32,33]. Nevertheless, we note that such kind of THz-spin-current-transmission experiments could be used to characterize any FM|NM interface involving metals. To conclude, the results of Fig. 4 support our THz emission model based on Eqs. (1) and (2) assuming a minor impact of the FM|NM interface.

*4.5. Extraction of the spin Hall angle*

As we have determined all relevant parameters in Eqs. (1) and (2), we can extract the relative magnitude of $\Theta_{SH}$. Figure 5 shows raw data comparing FM|W and FM|$Cu_{80}Ir_{20}$ heterostructures with a respective Pt-capped reference sample having an identical sample architecture (see Table 1). The absorptance of these references is 0.60 for $Co_{40}Fe_{40}B_{20}$(3 nm)|Pt(3 nm) and 0.61 for Fe(3 nm)|Pt(4 nm)|$AlO_x$(3 nm).

The ratios $\Theta_{SH}^{raw}$ of the bare THz-signal amplitudes (RMS) of sample and Pt-capped reference are also displayed by Fig. 5. As shown in our previous work[19], $\Theta_{SH}^{raw}$ can already provide a reasonable estimate of the relative spin Hall angle $\Theta_{SH}$. However, as shown by Eq. (2), this notion holds only true if the studied materials have comparable properties in terms of pump-pulse absorptance, THz conductivity and THz spin-current relaxation length (see Tables 2 and 3). Our detailed analysis finally allows us to fully account for these material parameters and to eventually extract the relative $\Theta_{SH}$. Here, Pt serves as the reference since it is by far the most studied spintronic material with $\Theta_{SH}^{Pt} = 12\%$ (Ref. 40).

Table 3 is the key result of this work: It summarizes the extracted values of $\lambda_{rel}$ and $\Theta_{SH}$ and compares them to the ratios $\Theta_{SH}^{raw}$ of the bare THz-signal amplitudes. When comparing to reported values of $\Theta_{SH}$, we find very good agreement for $Cu_{80}Ir_{20}$ (2.1% in Ref. 38). For W, the conductivity (see Table 2) indicates an α crystal phase for which a $\Theta_{SH}$ of similar magnitude and sign has been reported[52,17].

## 5. Discussion

Although the ratios $\Theta_{SH}^{raw}$ of the raw THz-emission signals in Fig. 5 already provide a reasonable estimate of the relative spin Hall angle (ref. 19), the direct comparison between $\Theta_{SH}$ and $\Theta_{SH}^{raw}$ (Table 2) clearly demonstrates the need for an advanced data analysis as provided by Eqs. (1) and (2). If only the raw data is considered, different optical and THz material properties as well as a varying $\lambda_{rel}$ may have a significant impact on the extracted $\Theta_{SH}$. For instance, variations by a factor of almost 2 occur in the case of $Cu_{80}Ir_{20}$ (see Table 3).

We find good agreement between the conductivities and spintronic parameters measured by THz spectroscopy here and values measured by contact-based DC electrical schemes previously. This observation and a recent study[21] of the THz spin-Seebeck effect suggest that highly excited electrons play a minor role in the THz emission process and that thermalized electrons make the dominant contribution.

As the analysis presented here is based on our model of spin transport [Eq. (1)] and THz emission [Eq. (2)], it is important to recapitulate its assumptions and limitations. First, Eq. (2) is only applicable for thin metal films (thickness below 20 nm), thereby ensuring a homogenous THz electric field into the depth of the metal (thin-film approximation). This condition is usually well fulfilled in practice. Second, in our analysis, THz emission by the anomalous Hall effect in the FM layer has been neglected, which, however, appears to lead to only minor corrections[26].

Finally, the spin-current transmission coefficients through the FM/NM interfaces are assumed to be approximately equal for all NM metals considered here. This assumption is supported by measurements using a Cu intermediate layer, but it is certainly important to investigate spin transmission through interfaces for a broader set of materials and interface parameters. Nevertheless, the results of this study (see Table 3) clearly show that THz emission spectroscopy is a reliable tool for contactless all-optical spintronic material characterization with high sample throughput.

## 6. Conclusion

We present a detailed measurement and analysis scheme to establish TES as a reliable, fast and efficient all-optical method to probe spin-to-charge conversion on the femtosecond time scale. TES has several advantages: Importantly, it allows for a large sample throughput because of the relatively short measurement time which is only about 1 minute for a complete THz waveform of the presented material combinations. Moreover, this all-optical and contact-free approach is very versatile as it does not require any microstructuring. It even permits the investigation of discontinuous films such as thin gold films below the percolation threshold[53]. We also present a promising experimental approach for FM/NM-interface characterization: THz-spin-current transmission measurements through a metal with negligibly small SHE. The good agreement of our measured spin Hall parameters with literature values suggests that thermalized electrons play a significant role in the THz-emission process. Finally, our results suggest copper-iridium alloys to be promising spin-to-charge-current-conversion materials. Future TES studies will also aim at quantifying the photon-to-spin conversion efficiency and the spin-current transmission through interfaces, eventually allowing for the determination of the absolute value of spin Hall angles.

**Acknowledgments**

We acknowledge funding by the the DFG priority program SPP 1538 "SpinCaT", the collaborative research centers SFB TRR 227 "Ultrafast spin dynamics" (project B02), SFB TRR 173 "Spin+X", the ERC H2020 CoG project TERAMAG/Grant No. 681917, the Graduate School of Excellence GSC266 "Materials Science in Mainz" and the H2020 FET-OPEN project ASPIN (grant no. 766566).